\documentclass[twocolumn,showpacs,prl,amsmath,amssymb,superscriptaddress]{revtex4}

\usepackage{graphicx}
\usepackage{dcolumn}
\usepackage{bm}

\begin{document}
\preprint{}

\title{Activated Transport in the individual Layers that form the $\nu_{T}$=1 Exciton
Condensate}

\author{R.D.~Wiersma}
\author{J.G.S.~Lok}
\author{S.~Kraus}
\author{W.~Dietsche}
\author{K.~von~Klitzing}
\affiliation{Max-Planck-Institut f\"ur Festk\"orperforschung, Heisenbergstrasse 1, 70569 Stuttgart,
Germany}
\author{D.~Schuh}
\author{M.~Bichler}
\affiliation{Walter-Schottky-Institut, Technische Universit\"at M\"unchen, 85748 Garching, Germany}
\author{H.-P.~Tranitz}
\author{W.~Wegscheider}
\affiliation{Fakult\"at f\"ur Physik, Universit\"at Regensburg, 93040 Regensburg, Germany}

\date{June 29, 2004} 

\begin{abstract}

We observe the total filling factor $\nu_{T}$=1 quantum Hall state in a
bilayer two-dimensional electron system with virtually no tunnelling. We find
thermally activated transport in the balanced system with a monotonic increase
of the activation energy with decreasing $d/\ell_B$ below 1.65. In the imbalanced
system we find activated transport in each of the layers separately, yet the
activation energies show a striking asymmetry around the balance point. This implies
that the gap to charge-excitations in the {\em individual} layers is substantially
different for positive and negative imbalance.

\end{abstract}


\maketitle

\par
The Bose-Einstein condensate (BEC) is an ordered state of a many
particle system with properties which do no longer depend upon the many
individual wave functions but rather upon a single macroscopic one. Presently
known BECs include superconductors, the two Helium isotopes and rarefied atomic
vapours. Excitons, consisting of a hole in the valance band bound to an electron
in the conduction band in a semiconductor, have long been suspected to form a
BEC as well~\cite{Rice}, but initially the short life times and the intrinsic
self-heating of optically generated excitons prevented condensation. More
recently, however, optically generated indirect excitons displayed
features arising from collective behaviour~\cite{Butov1,Butov2,Butov3,Chemla}
\par
Indirect excitons can be produced in spatially separated quantum wells where
they have infinite life times and long mean free paths and where they can be
cooled down to lowest possible temperatures. Recently, it has been shown that a
BEC most likely exists in double quantum wells (DQW) where each of the two
wells contains a two-dimensional electron gas (2DEG) at a half filled Landau
level in the appropriate magnetic field~\cite{Kellogg1}. In this system, the
excitons are formed by the pairing of empty and filled electron states in the
conduction band in the two layers.  In drag experiments, where current is
passed through one of the layers (''drive'' layer) and the induced voltage
drop in the other (''drag'') layer is measured, the signatures of the new
state can be seen: an activation gap in the longitudinal resistance and a
quantised Hall drag which is identical to the Hall voltage of the current
carrying layer~\cite{Kellogg1}. As a consequence, when identical but
counter-flowing currents are passed through the two layers then a dramatic
vanishing of both the longitudinal and the Hall resistance is
observed~\cite{Kellogg2}. More recently, similar phenomena have also been
observed in coupled 2D hole gases~\cite{Tutuc1}.
\par
The requirement for observing this novel superfluid state is the ratio of the
interlayer Coulomb interactions (parameterised by the distance ($d$) between
the 2DEGs) and the intra-layer Coulomb interactions (parameterised by the
magnetic length ($\ell_B \equiv \sqrt{\hbar/eB} = 1/\sqrt{2\pi n_{T}}$) at the
half-filled Landau levels) being below a critical value of about 1.8~\cite{Kellogg1,Kellogg3}.
Here $B$ is the magnetic field and $n_{T}$ the total electron density of the bilayer.
$d/\ell_B <$ 1.8 requires DQWs containing two 2DEGs with respective densities of about
$3\times$10$^{14}$ m$^{-2}$ separated by barriers of $5$ to $20$ nm thickness.
Mobilities exceeding $40$ m$^{2}/$Vs at these densities are required in order
to prevent electrons from becoming localised before the required magnetic field
is reached.
\par
In this Letter we reveal unexpected properties of the novel superfluid condensed
exciton state. We study electric transport in the condensed phase, find a
thermally activated behaviour, and determine the dependence of the activation
energy on the coupling parameter $d/\ell_B$. Strikingly, upon producing a symmetric
density imbalance at a constant total filling factor 1 ($\nu_T$=1), we observe a huge
asymmetry of the activated transport in each of the individual layers.
This implies that the measured activation energy is not connected with the condensation
energy of superfluid $\nu_{T}$=1 state. Instead it reflects the gap to charge-excitations
of the individual layers which turns out to be substantially different for positive and
negative imbalances. Our data additionally demonstrate that
measuring both layers in parallel easily leads to erroneous conclusions~\cite{Tutuc2}.
\par
Our DQWs consist of two $17$ nm GaAs quantum wells separated by a superlattice
of $12.4$ nm total thickness made up of alternating 4 monolayers (ML) AlAs
($1.13$ nm) and 1  ML GaAs ($0.28$ nm). The electrons originate from bulk
doping with Si that is placed 300 nm below and 280 nm above the wells,
respectively. The two 2DEGs have intrinsic densities ($n_U,n_L)$ of $\sim 4
\times 10^{14}$ m$^{-2}$ in both the upper and the lower layer and mobilities of $70$
m$^{2}/$Vs. Eight Ohmic contacts were made to the upper 2DEG and six to the lower 2DEG using
metallic frontgates~\cite{Eisenstein2} and buried backgates~\cite{Rubel} for contact separation.
The densities in the layers can be adjusted independently by another front- and backgate
atop and below the Hall bar that has a width of 80 $\mu$m and a length of 900 $\mu$m.
Interlayer leakage is negligibly small: at 50~mK and in zero magnetic field, the
interlayer resistance is several G$\Omega$ and the dI/dV shows no resonant tunnelling
peak within the noise level (0.5 $\times$ 10$^{-9}$ $\Omega^{-1}$). From the observed zero
bias tunnelling peak at $\nu_T$=1 (the dI/dV is about 9 $\times$ 10$^{-8}$ $\Omega^{-1}$
with a width of 10 $\mu$V), we deduce that tunnelling is at least 50 times smaller
than in previous experiments~\cite{Spielman2}. Transport measurements were done
in a dilution refrigerator at temperatures down to 35~mK. AC currents of
0.1-0.5~nA at 1.2 to 6 Hz were used to measure the longitudinal and Hall
resistances. The independent contacts to the layers additionally allowed
measurements of the drag, i.e. the voltage in the ''drag'' layer divided by
the current in the ''drive'' layer. The linearity of the measurements
was tested under all measurement conditions and no significant deviations were
found up to $\sim1.0$ nA.

\begin{figure}[t]
\centering
\includegraphics[width=80mm]{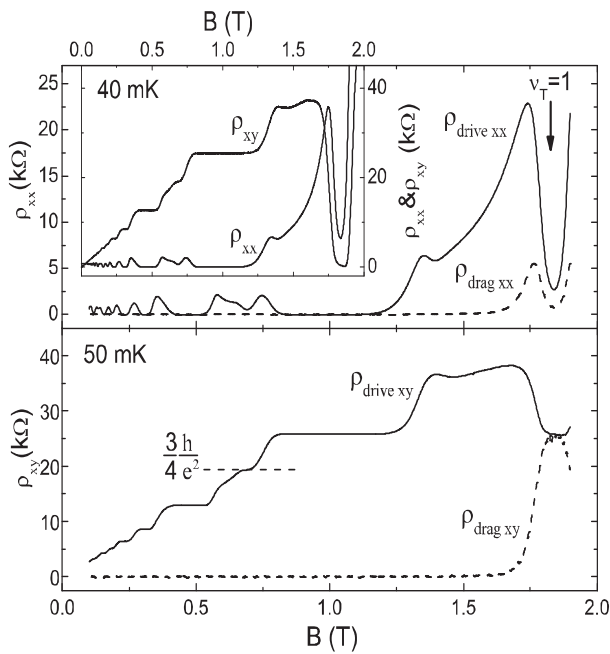}
\caption{(top) $\rho_{drive,xx}$ and $\rho_{drag,xx}$ and (bottom)
$\rho_{drive,xy}$ and $\rho_{drag,xy}$ vs. magnetic field at 50~mK and matched
densities ($n_U$=$n_L$=2.22$\times$10$^{14}$ m$^{-2}$ corresponding to
d/$\ell_B$=1.57). At $\nu_{T}$=1 both longitudinal components tend to zero,
while both Hall components tend to $h$/$e^2$. Inset plots the counterflow
experiment; $\rho_{xx}$ and $\rho_{xy}$ both tend to zero.
}
\label{fig1}
\end{figure}

\par
Data taken with a density in each layer of $2.22 \times 10^{14}$ m$^{-2}$ are
shown in Fig.~\ref{fig1} that plots longitudinal ($\rho_{drive,xx}$) and Hall
($\rho_{drive,xy}$) resistances of the layer to which the current is applied,
as a function of magnetic field. At $\nu_{T}$=1, $\rho_{drive,xx}$ shows a
pronounced minimum while $\rho_{drive,xy}$ drops to approximately 25.8
k$\Omega$ ($h$/$e^2$). Away from $\nu_{T}$=1, the traces show Shubnikov-de Haas
oscillations of a single layer. At temperatures above $\sim$ 250 mK, the
minimum in $\rho_{drive,xx}$ and the quantisation of $\rho_{drive,xy}$ have
disappeared. Also shown are the longitudinal drag ($\rho_{drag,xx}$) and the
Hall drag ($\rho_{drag,xy}$). The approximate quantisation to $h$/$e^2$ in
$\rho_{drag,xy}$ indicates the formation of the superfluid exciton condensate.
This is verified by sending equal but counter-flowing currents through each of
the layers simultaneously. Indeed as recently observed~\cite{Kellogg2,Tutuc1},
both the longitudinal and the Hall voltages in the layers tend to zero at the
lowest experimental temperatures (inset Fig.~\ref{fig1}). 

\begin{figure}[t]
\centering
\includegraphics[width=80mm]{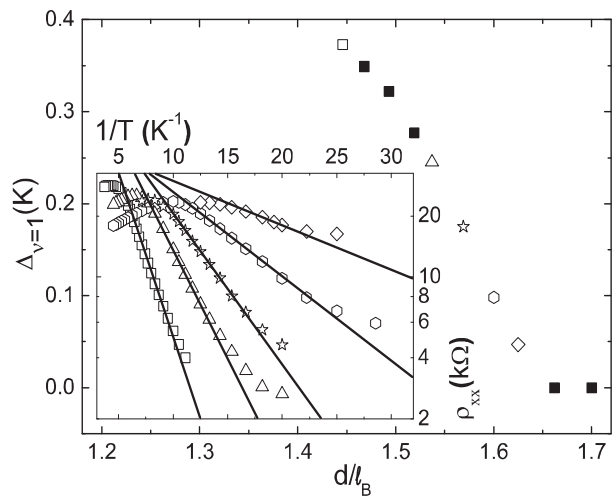}
\caption{Activation energies of the $\nu_{T}=1$ state obtained from
$\rho_{drive,xx}$ with balanced layer densities for various d/$\ell_B$. The inset
shows Arrhenius plots of some of the data with corresponding symbols in the
main panel; lines are fits to $\rho_{xx} \propto$ exp(-$\Delta_{\nu=1}$/k$_B$T).
}
\label{fig2}
\end{figure}

\par
We now turn to the temperature dependence of the longitudinal resistance at
$\nu_{T}$=1. The inset of Fig.~\ref{fig2} plots $\rho_{drive,xx}$ of the
lower layer vs. the inverse temperature for a series of different $d/\ell_B$
values, obtained by adjusting both the front- and backgate. In all cases one
can distinguish between three different temperature regimes: at higher
temperatures the resistance is only weakly temperature dependent as expected
for filling factor 1/2 for a single layer. With decreasing temperature there is
an exponential decrease of the resistance which, in the end, levels off into
saturation around $\sim$50~mK. At present it is not clear if this is an
intrinsic phenomenon or caused by insufficient cooling of the 2D electronic
system below 50~mK. From the intermediate exponential range we deduce
activation energies ($\Delta_{\nu=1}$) which are plotted in the main panel of
Fig.~\ref{fig2}; the symbols in the inset correspond to the symbols in the main
panel. Also included are zeroes corresponding to measurements that did not
display a minimum at $\nu_{T}$=1 at the lowest temperature.
\par
The activation energy shows a monotonous increase with decreasing $d/\ell_B$ below
a certain $d/\ell_{B,crit}$ which is $1.65$ for our sample. This $d/\ell_{B,crit}$ is
significantly smaller than the value of $\sim$1.83 reported previously~\cite{Kellogg1,Kellogg3}
and it could possibly be due to our slightly lower mobility or our much smaller interlayer tunnelling.
The increase of $\Delta_{\nu=1}$ below $d/\ell_{B,crit}$ reminds of the behaviour at a phase
transition, for example that of a gap in a superconductor. It is, however, not
clear what type of excitations are contributing to the measured
$\Delta_{\nu=1}$. In particular, it is not clear if $\Delta_{\nu=1}$ is
directly related to the condensation energy of the excitonic state or if it is
even a property of the collective bilayer system at all. Naively thinking, one
expects to measure zero resistance in a superfluid as soon as the temperature
is lower than the superfluid transition temperature. The fact that in our
experiment the resistance does not abruptly drop to zero then suggests that at
finite $T$, excitations are present in the system that cause a finite
resistance. Presumably it takes a finite energy ($\Delta_{\nu=1}$) to produce
such excitations. The observation that for different $d/\ell_B$ the exponential
decrease of the resistance saturates at somewhat different temperatures could
hint that even at the lowest $T$, some of these excitations remain present.
In this context, it has been pointed out recently~\cite{Fertig,Kellogg2,Eisenstein}
that the finite longitudinal resistance could originate from a current-driven
flow of vortices across the layers, similar to the case of certain type-II
superconductors~\cite{foot1}. Although in particular our observed linear
current-voltage characteristics (as in ~\cite{Kellogg2}), do not agree with the
theory~\cite{Fertig}, our measurements of the temperature dependence of
$\rho_{xx}$ of the imbalanced system prove that the measured
$\Delta_{\nu=1}$ is not related to the condensation energy of the {\em
total} system. Instead we deduce that it must be connected to excitations
of the {\em individual} layers.

\begin{figure}[t]
\centering
\includegraphics[width=80mm]{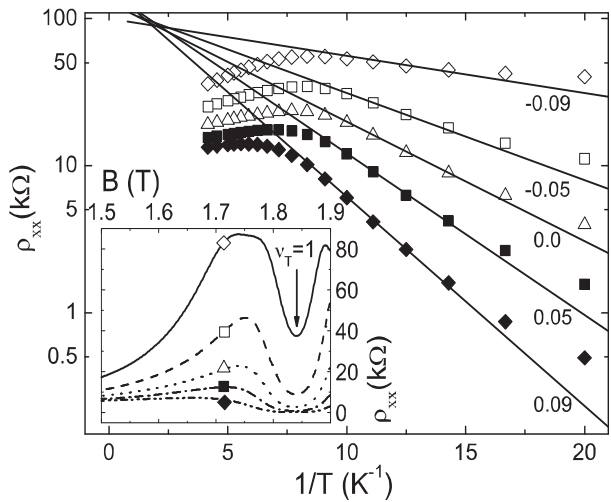}
\caption{Arrhenius plot of the lower layer $\rho_{drive,xx}$ for 5 different
imbalances ($\equiv$ [$n_L$-$n_U$]/$n_{T}$) indicated in the right of the
figure. The inset plots the lower layer $\rho_{drive,xx}$ at 50~mK vs. magnetic
field for these 5 imbalances. The total electron density is fixed at $4.44
\times 10^{14}$ m$^{-2}$ corresponding to $d/\ell_B$=1.57.
}
\label{fig3}
\end{figure}

\par
Below we study the transport in each of the layers separately at imbalanced
electron densities, yet at a constant total electron density. The front- and
backgates were adjusted to have a total density of $2n=4.44 \times$ 10$^{14}$
m$^{-2}$ equally distributed between the layers, corresponding to
$d/\ell_B$=1.57 ($\star$ symbol in Fig~\ref{fig2}). Then interlayer bias was added
that produces a symmetric imbalance between the two layers, i.e. one layer
had $n+\Delta n$, while the other had $n-\Delta n$. Next, $\rho_{drive,xx}$
and $\rho_{drag,xx}$ where measured as a function of temperature. Then the
drag and drive layer were interchanged and the procedure was repeated. We
have done measurements for several imbalances
($\equiv$[$n_L$-$n_U$]/$n_{T}$) between $-0.1$ and $+0.1$. To check for
consistency, for two of the measurement points, the front- and backgates were
fine-tuned to exactly produce the symmetric density imbalance and no interlayer
bias was used. In these cases, identical results were obtained.

\begin{figure}[t]
\centering
\includegraphics[width=80mm]{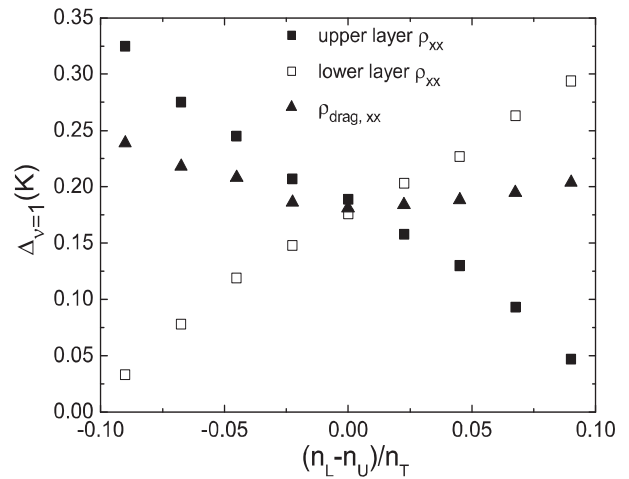}
\caption{Activation energies of the $\nu_{T}=1$ state vs. layer imbalance.
($\square$) and ($\blacksquare$) correspond to the activation energy determined
from the longitudinal resistance in the lower and upper layer, respectively.
($\blacktriangle$) denotes the activation energy obtained from the
longitudinal drag. 
}
\label{fig4}
\end{figure}

\par
Throughout the range of density imbalances studied, the Hall drag remained approximately
quantised at the lowest temperature, yet the temperature dependence of both longitudinal
resistance and longitudinal drag changed significantly. Typical data are shown
in Fig.~\ref{fig3} that plots the lower layer $\rho_{drive,xx}$ vs. inverse
temperature for various density imbalances indicated in the right part of the
figure. Strikingly, upon increasing the imbalance from negative to positive values
(i.e. increasing the lower layer density), the activation energy increases.
Furthermore, when interchanging the role of the two layers and sending
a current in the upper layer, we find that the upper layer $\rho_{xx}$ at a given
imbalance resembles very closely the lower layer $\rho_{xx}$ for minus that
imbalance. As a consequence, the activation energy determined from the upper
layer $\rho_{xx}$ decreases with increasing imbalance (i.e. it increases with
increasing the upper layer density). This asymmetry of the activation energies
of the individual layers with imbalance is summarised in Fig.~\ref{fig4}. It proves
that the measured activation energies are not directly connected with the condensation energy
of the total system that should be the same, regardless of which layer has the lower
density and which the higher density. Instead, it implies that the activation energy
reflects the gap to charge-excitations in the {\em individual} layers and that the excitation
spectrum in the individual layers is substantially different for positive and negative
imbalance.
\par
Our non-symmetric behaviour of the activation energies around the balanced density
point seems to contrast sharply to previous measurements in hole bilayers~\cite{Tutuc2,Hamilton}
that found a symmetric behaviour around balanced densities. Both of those experiments however,
measured the two layers {\em in parallel} (i.e. no separate contacts to
the individual layers existed). We note that the resistances of the individual layers
in the slightly imbalanced system are extremely different.
It is thus evident that in the imbalanced case, most of the current flows in the higher
density, less resistive layer and that mainly the properties of this layer are
probed. Indeed, when we measure both layers in parallel, but also when we study the activated
longitudinal drag that probes the coupled system ($\blacktriangle$ symbols in Fig.~\ref{fig4}),
a more symmetric behaviour is observed.
\par
The very different resistances of the individual layers of the slightly imbalanced
$\nu_T$=1-state also shed new light on the observed disappearance~\cite{Tutuc2} of the
insulating phase for filling factors slightly larger than 1 with imbalance. In particular,
it strongly questions its interpretation in terms of a pinned {\em bilayer} Wigner
crystal~\cite{Tutuc2}. The inset of Fig.~\ref{fig3} plots $\rho_{xx}$ of the lower
layer at 50~mK vs. magnetic field for the imbalances indicated in the main figure
with corresponding symbols. It now becomes evident that upon decreasing the lower layer density
(traces marked with $\square$ and $\lozenge$ symbols) the insulating phase in the lower layer
becomes much stronger, while simultaneously the upper layer gets a higher density and its
insulating phase disappears (traces closely resemble those marked with $\blacksquare$ and
$\blacklozenge$ symbols). Consequently when measuring the layers in parallel, almost all
current flows in the upper, less resistive layer and the acclaimed bilayer
Wigner crystal~\cite{Tutuc2} seems to disappear. Our data show however, that the insulating
phase actually survives in the lower-density layer. We further note that this insulating
behaviour in the lower-density layer is not simply due to its somewhat reduced mobility,
since reducing the total density such that both layers have this lower density, results
in a much weaker insulating phase than that observed in the lower-density layer in
the imbalanced case. 
\par
Finally we note that for $d/\ell_B$ slightly higher than the critical value (where
no $\nu_T$=1 state can be observed for matched densities), a density imbalance can induce
the $\nu_T$=1 state. A similar observation was made previously in coupled 2DHGs~\cite{Hamilton},
and very recently also in coupled 2DEGs~\cite{Spielman}. Under such circumstances, we
observe a minimum in $\rho_{xx}$ at $\nu_T$=1 with thermally activated behaviour only
in the higher-density layer, while $\rho_{xx}$ of the lower-density layer shows no trace
of the correlated state at all.
\par
Summarising, we have determined activation energies for transport in the balanced
$\nu_T$=1 state over a wide range of the coupling parameter $d/\ell_B$ and
find a monotonous increase with increasing coupling below 1.65. In the symmetrically
imbalanced $\nu_T$=1 state an asymmetry in the activation energies of the longitudinal
resistances of the {\em individual} layers was observed. In each layer, this activation
energy increases approximately linearly with increasing the density of the respective layer.
This proves that the measured activation energies are not connected with the condensation
energy of the total system. Instead it implies that they reflect the gap to charge excitations
in the {\em individual} layers building up the $\nu_T$=1 exciton condensate and that the
excitation spectrum of the individual layers is rather different for positive and negative
density imbalance. 
\par
We thank R. Gerhardts and W.~Metzner for critical reading of the manuscript and acknowledge
the BMBF for financial support (01BM913/0).

\end{document}